# Encounters with Alex
by
Kristian Fossheim
NTNU, Trondheim

In the 80th anniversary book for Alex Müller I wrote a story of our scientific collaboration, *Shared Fascinations*. This time I will be more personal, about the human side of our collaboration and encounters, while also referring to episodes mentioned in *Shared Fascinations.*

**Early days**
Our collaboration started in 1968. I had returned to Oslo from University of Maryland in the fall of 1967 after two years there, first on a NATO Fellowship, then as an appointed research associate. I had become totally fascinated by superconductivity, and had in particular been captivated by the sudden change of properties at the transition temperature. I had observed, and described theoretically, the transition of submillikelvin sharpness in metals.  Alex and I had never met, but we now agreed, thanks to contact through our common friend Jens Feder, to collaborate in the study of another type of solid state transition, the structural change occurring in SrTiO$_3$.

We collaborated via telephone calls between University of Oslo and the IBM lab in Rüschlikon, near Zurich. Based on results from ultrasonic measurements, we soon agreed to write a paper for PRL on critical attenuation of high frequency acoustic waves near the structural phase transition in SrTiO$_3$ at T$_c$= 125 K, in a collaboration with my old physics mate from student days, Bjørn Berre. I always made sure IBM got the telephone bill since telephone calls abroad were fiercely expensive in those days. One colleague in Oslo almost got his phone confiscated because he made so many long international calls. Our paper was sort of a breakthrough since it seemed to show that the behavior in SrTiO$_3$ was dominated by order parameter fluctuations, hence not classical, or Landau-like, while at the same time Alex had shown classical temperature exponent for the order parameter on approaching T$_c$. To investigated this further, Bjørn and I did additional measurements of elastic properties by means of sound velocity. Instead of a smooth transition from one level to another across the transition, we found a minimum, thus indicating non-classical, non – mean field like behavior.

Now it was time, in 1970, to go to a meeting on phase transitions in Varenna, at Lago di Como, from Trondheim, where I had just moved. On the way I flew via Zurich, half-way planning to visit Alex at the Rüschlikon lab. Somehow, I was so much in awe of this man whom I had never met, only spoken with on the phone, that I continued to Varenna without visiting the lab. I am sure he could never suspect this, and I never told him the reason. When we finally met in Varenna, I saw a strong- minded black- haired, slightly moustached sharp-looking guy, confirming my impression that this was a very determined and intelligent man, who deserved my greatest respect. Even though I had just spent two years among great physicists at University of Maryland, like Richard Ferrell and Rolf Glover, I thought this was a really tough one. He was only nine years older than I, but this represented a huge gap in knowledge and experience in my mind, not least because I discovered he spoke both French and Italian fluently, in addition to English, German and Swiss- German dialect, with a rasping r. In addition he was a true champion of EPR, which was outside my experience at the time.

However, in spite of everything, it still turned out I had something important to teach him: The velocity minimum of sound at $T_c$ in $SrTiO_3$, indicating non-classical behavior, thus contradicting Alex´s measurements of the order parameter. He became very interested, and immediately suggested that Bjørn and I should write a PRL. I thought the data were not quite up to PRL standard, and did not follow the advice. But Alex went straight back to Zurich after the meeting, measured much more carefully close to the transition, and found an exponent of about 0.3 instead of 0.5 very close to $T_c$. I believe this was the first instance where a non- classical exponent was measured at a structural phase transition. Alex, on the other hand, soon published a PRL on his new data, and the matter was settled.

I guess Alex was pleased with this outcome, and I got the message from another physicist at IBM that I was being considered for a job there. Eventually, instead, I was invited for two summer stays, in 1973 and 1975, when I worked on totally different matters with Ulrich Høchli. During conversations with Alex I understood he was much aware of my experience in superconductivity from Maryland and Oslo, where I had several PRLs, and other papers. Alex told me he never really worked in that field. I got the impression he half way envied me this experience. We all know this would change dramatically later on!

I also discovered that Alex was an automobile *aficionado* and had a Plymouth Valiant, just like me, except I had a newer model, 1966. I understood he was his own mechanic, and shared this interest with his colleague Heini Rohrer who also had a Plymouth Valiant. In addition, as it would later turn out, Alex had several cars, including a 12 cylinder Rolls Roys, in which I was invited many years later to ride with him and Inge to a concert in Zurich.

**Some differences**

During my sabbatical at IBM Thomas J Watson Research Center 1975-76 one of my main goals was to get involved with the newly developed phase conjugation techniques in the study of solids. I worked there with Norm Shiren, Bob Melcher and Koji Kajimura. Again, Alex`s name came up. He made us aware of work at the Ioffe Institute in Leningrad, where even powders showed echo-like phenomena in the spirit of spin echo. The three of us set out to work on it, and it became a major project. Norm Shiren had established the technique in so- called echo single crystal materials which showed this highly nonlinear effect of "time reversal". I came upon the idea to use it as a tool, where the echo material would work like a very advanced time reversal transducer, and could be used to study another material which could be attached. I recruited a PhD student, Rune Holt, in Trondheim, to use this method when studying critical phenomena at phase transitions where any small inhomogeneities would distort the signals. This was a very successful research project, and again the phase transition in perovskite materials was the object.

In this connection a misunderstanding happened, where Alex referred to our method as invented by Shiren. I protested, and there was a somewhat heated exchange between us. I believe I was able to explain the real situation, eventually.

In 1983 I invited Alex to Trondheim to smooth out things for one more reason: One of my PhD students, Sigmund Stokka, a very talented, strongminded fine guy had been doing a nice job to reveal the predicted crossover behavior in perovskites like $SrTiO_3$. By specific heat measurements the phase boundary between cubic and tetragonal structure under pressure was determined, and led to determination of the corresponding crossover exponent of 1.25, for the first time, in $SrTiO_3$, a project Alex had had on his wish list, but not yet accomplished. Sigmund and I had discussed the reference list in the ensuing paper, and there was a reference to Alex´s work I wanted to include. Sigmund disagreed, and I said "This will give me trouble with Alex." Sure enough, it did. Alex let me know. Now, Sigmund had left for a job in Stavanger. What could I do? I invited both of them to Trondheim, and the issue was brought up in a meeting between the three of us in my office. Alex was not in a good mood, and Sigmund stubbornly argued his case. There was a stalemate. Then I told Alex: "You have to let young people do some mistakes." Alex immediately softened and agreed. The case was closed right there.

I offered Alex to take my Valiant and look around Mid- Norway with Inge, which they enjoyed. Next, he came along to our cabin on the coast together with my wife, Elsa, and our son Terje, while Inge took the Costal Steamer to Bergen to experience the famous mountain railway to Flåm, and the magnificent Sognefjord. Meanwhile, at the cabin, we experienced a great storm in Taftøysund, and our boat, which was moored in the harbor, was in trouble. Alex, Terje and I climbed down a very steep rock to the water and saved the boat. An operation Alex enjoyed. The cabin was primitive style where everything had to be carried in and carried out, and Alex took his turn helping Elsa in the kitchen. The cabin was quite new, and Alex was our first guest there. Regrettably, we did not even yet have a guest book for the record of visits, and surely we did not anticipate the fame our first guest would reach. Alex often later referred to the visit and the rescue operation as a nice experience.

Later, Alex continued the crossover experiments in his own lab, and got precisely similar results as we had, but in a different material, $LaAlO_3$. I have to admit I was proud to have been first. Apparently, the PRL referee was not equally impressed, and did not want to follow the story of this new exponent in the saga of critical phenomena, so Phys. Rev would have to do.

### Stockholm

When the news of the discovery of cuprate superconductors came out, I was astonished at the fact that exactly those insulating materials we had collaborated on were at the center of the discovery. It was already late, formally speaking, for nominations for the Nobel prize in physics. But I had regularly received invitations to nominate. I had already nominated Binnig and Rohrer for the 1986- prize, and I surely did not miss the opportunity this time, although I am afraid it was a close call. Very briefly I stated that the discovery was so obviously worthy of the prize, that the world would not understand it if this opportunity was missed. On the day of official announcement I was sitting in my office counting down minutes before the official broadcast from Stockholm. It was full score! Bednorz and Müller got it. My fax to IBM Rüschlikon, congratulating the winners, was sent just minutes later, and apparently it was one of the first they received, maybe the very first.

On approaching December 10, 1987, I received official letter from the Nobel Institute that my wife and I were invited for the event, and hotel reservations were made at Grand Hotel where all the laureates and their families were to stay. So we had been invited on the laureates´ quota by Alex. We flew over and moved in, and found a printed program on our hotel bed where, to our surprise, our modest presence was included. Grand Hotel in Stockholm has a dignity and style fit for the occasion.

In the evening, walking out of the hotel door for a minute to draw some fresh Stockholm air, I spotted a tall man standing there looking at the hotel facade. I recognized him, walked up to him and greeted him. It was Heini Rohrer, recipient of the physics prize the year before, shared with Gerd Binnig, also from the IBM lab in Rüschlikon, in itself an unbelievable and unprecedented occurrence. I asked Heini why he was standing there instead of going inside. He said he wanted to recall the experience from the year before and would not enter Grand Hotel again. He wanted to cherish and protect his memory from that time.

We were treated practically like family member. During the award ceremony in the Concert Hall Elsa and I were given seats up front with the laureates` families, with Inge. Elsa was of course dressed in the finest outfit she ever wore, and to her great satisfaction, her green dress matched perfectly with that of queen Silvia!

The ensuing banquet in the City Hall was out of this world. A procession came along the balcony and down to the hall, headed by King Carl Gustav and Inge. The food was just exquisite, and had been in preparation for months. I believe Alex gave the after dinner speech, but I am sorry to say I have forgotten the contents. After dinner there was dance in the fabulous room above and aside from the main hall, and we met Alex´s father and could see him dance there, very much able. Alex was himself close to 60, so his father must have been well into his 80´s. We had the pleasure to join the dance before retreating at midnight. During the night we met several other great friends from the superconductivity science scene. In a very happy photo shot I still have, but this one from my next visit to the Nobel ceremony in 1992, there are Alex and Inge, Øystein and Hanna Fischer from Geneva, and Mrs Kitazawa from Tokyo, my wife and I, the shot having been taken by Koichi Kitazawa. Of these, two great men and friends, Øystein and Koichi both have left us in recent years, to our grief and sorrow.

**Trondheim again**

The life of a Nobel laureate is a whirlwind, before as well as after the event. Personally, I had long ago booked Alex for a visit to Trondheim, directly after Stockholm. So he came with Inge. I received them at the airport with the official limousine of the university, riding back to Trondheim with our guests, joined also by our rector. Of course, our guests were installed at Trondheim`s finest hotel, Britannia. Later, the same evening we took Alex and Inge to our house for dinner where Elsa served a salmon dish, and I had chosen a German *spätlese* to go with it. As soon as we had finished, I offered Alex my new and comfortable armchair, which Elsa had given me for my 50th birthday. Alex quickly fell asleep. Finally, he could relax completely among friends! He needed it badly.

However, next day was another busy day with interviews by several media, and a lecture before a packed audience of 250 or more, meeting with my group, and finally a well prepared banquet at the university´s representation building, the Lerchendal Mansion, with leading persons from the university and from IMB Norway who sponsored the event. In my speech during dinner I said that when I was collaborating with Alex in the 70`s I often thought that this physicist was of a caliber which would have earned him the Nobel prize, had he worked in a popular field like superconductivity. But he was not there at the time. Well, eventually he did come to superconductivity, and here he was, having won the prize.

For entertainment I had invited our son Øyvind and his talented music student friend Cecilia. Inge remarked that she could see a future as a really fine violinist in our son, but added that it would be very costly for him to prepare for an international career. Of course,

Inge knew the world of music, herself having been an established soprano. At that time I thought wrongly that music would be like physics, just a matter of following the usual university path and work hard. I have later learned that the area of music is very different, and much tougher. The performance brought back memories from the summer of 1973 when I was at the IBM lab with Alex, and we visited their home in a small village behind the Albis, with our children, Kristin 10, Terje 6 and Øyvind 4. At the barbeque outside their house, Alex was charmed by the small boy Øyvind. He gave him a half dollar coin and commented: "He looks you in your eyes, and you melt!" Well, here at the Lerchendal Mansion was Øyvind again, now an 18 years old, young man. I recalled then, that it was the 16th of December, Beethoven´s birthday, and also the date of my thesis defence at University of Oslo many years back. After Øyvind and Cecilia had played, I followed them to the car and waived them off. After that evening they were fiancées. They married some years later, and became parents of two great boys: Andreas, now an ambitious physics student at my university, following in the footsteps of his grandfather it seems, and Yngve, a very talented chess- and football player, dreaming to be a football star. Both Øyvind and Cecilia were soon to become permanent members of the Oslo Philharmonic Orchestra. But as it often happens, have established new families. Alex has since mentioned this particular visit to Trondheim on several occasions, as a nice experience.

**Two on the roof, in Pisa**

Of the many episodes and encounters I have had with Alex, perhaps one stands out specially. In the aftermath of the discovery of the high-$T_c$ superconductor YBCO I specially wanted to do two things: 1) Measure the specific heat anomaly at $T_c$, looking for non- BCS like behavior due to the low concentration of carriers, and 2) measure the elastic behavior for lattice softening related to the high $T_c$. Alex and George Bednorz immediately provided the samples we needed. At the first European high-$T_c$ meeting in Pisa in the spring of 1987 I could report quite unusual elastic behavior: a strong softening of the lattice on cooling to low temperatures. This was exciting, since it might be related to cooper-pair creation at unusually high temperature. Times were very busy, and on the plane from Copenhagen to Milano I finally wrote a handwritten version of a paper with Alex and George, accompanied by a taste of *Chianti classico.* On arriving in Genova for the meeting, Alex was in a constant whirlwind of people who wanted his attention. How could we have a few moments alone to discuss the paper? We found a stairway leading up to the roof. We got out there and sat down in the sunshine. I handed Alex my handwritten note. He read through, gave it back to me, and said "OK, publish." The paper soon appeared in Solid Sate Communications, and was very regularly cited. It also turned out I had been right about the non- classical specific heat. In my group we measured a broad transition, quite different from the BCS picture which has an extremely sharp onset of the transition coming from above in temperature. In low-$T_c$ classical superconductors I had already at Maryland shown that the transition could be defined with a sharpness of sub – millikelvin, as mentioned. This was indeed different, more like a thousand times broader, absolutely not BCS- like! The problem was, literally nobody else seems to have any clue about the fundamental difference between this anomaly and that of usual low-$T_c$. Everybody analyzed the anomaly as BCS- like. It took many, many years before people in the field understood, or accepted, that this was a fundamentally different case, that vortex fluctuations played a major role, as Asle Sudbø and his students in Trondheim calculated convincingly.

**Alex in Trondheim again: Doctor Honoris Causa**

In 1992, the board of the University of Trondheim had put me in charge of a new strategic research office, called the Research Academy of the University of Trondheim. Alex was selected that year as the university´s only candidate to be conferred the title of Doctor Honoris Causa. It was awarded in a major ceremony in which all new PhDs of the entire university that year were given their diploma. It was, in fact, the first such event, and my responsibility, and I had chosen the great rotunda of the Student Union building for the ceremony. I gave the required ceremonial speech for Alex, as well as the after dinner speech at the banquet in the evening. In that speech in 1992 I talked about future challenges in science, and mentioned three special examples, quote:

*-1. Understanding the human brain is a science yet in its infancy. Gaining deeper insights into this marvelous instrument belongs to the greatest challenges in science today. What can we expect the future will bring?*

*-2. The genomes of life forms show broad variations which we are just beginning to map out. This is an urgent task, since, as it is said sometimes: "The library of life is on fire". The destruction of life forms must be stopped! And genetic science must be kept on a constructive course.*

*-3. Chemistry and physics offer infinite possibilities for new substances and materials, with, as yet, unimagined consequences for science and technology.*

It is tempting to comment on these points today: Regarding point 1 above, the later Nobel laureates May-Britt and Edvard Moser who received the Nobel prize for their brain research at our university in 2015, were at that time, 1992, students in Oslo. Regarding point 2., the human genome was unraveled some 10 years later. And I have myself taken the initiative, and conducted the award process of the first major global prize for sustainability science, in Trondheim, The Gunnerus Sustainability Award, while president of the Royal Norwegian Society of Sciences and Letters, first time in 2012. Regarding point 3, it is especially interesting to point out the later huge developments in nanoscience.

As a kind of amusing *a propos*, between the ceremony and the banquet, I took Alex on foot to the center of the city to inspect an outdoor display of American veteran cars. He made his expert comments and enjoyed the display. The next day, I gathered my PhD students and their families and Alex for a two hour hike in Bymarka, the wide outdoor area and hills west of Trondheim, Alex wearing my old checkered sweater, as photos from the tour show. I still keep it in my closet, known as it is in the family as the "Müller sweater."

**Nobel book and follow up**

In 2004 I published a textbook on superconductivity and its applications with my colleague Asle Sudbø. In preparation for this, I traveled along the East Coast of the US, and in Europe, to interview Nobel laureates for a chapter on personalities who have contributed most to the development of the field. Later, these interviews were printed in full in a book on Springer in 2013: *Discoveries and Discoverers. Ten Nobel Laureates Tell Their Story.* My interview with Alex took place in his office at University of Zurich, and he told vividly about his life, his venture into the field of superconductivity at the age of fifty, like a graduate student he said, during a stay at IBM Thomas J Watson Research Center in New York. And later experiences were covered, of course.

My book was published almost ten years later. Before that, in the winter of 2012, I visited Alex and Inge who had moved to their *Tertianum*, a Swiss term for retirement homes, where he still enjoyed working in his private office, and where diplomas and reminiscences from all his 22 honorary degrees could be spotted. The whole place was in fact a very nice setting, fitting his status. As a kind of an ironic case, among the inhabitants of the place, Alex pointed out a small, very old man who, he said, had been a German general during World War 2. I was invited to lunch with Inge and Alex in the very nice restaurant of the place. They were both in a good mood, and later Alex and I discussed both history and the contents of the interview. As usual, Alex, at 85, had done his homework, and was as keen on physics as ever, showing his most recent publications.

Since then I learned that my visit had been much appreciated, and our contacts continue. Old friendship has not faded away, and as far as Alex`s stamina in physics is concerned: No end in sight!

My encounters with Alex have happened in numerous countries and places all over the Northern Hemisphere. He has been my greatest inspiration in physics, and through this has enriched my life and that of my entire family.

Just as I am writing these lines, the Nobel prizes for physics 2016 are announced. Again, phase transitions of matter are the focus. It reminds me that both Alex and I have had the privilege to participate in, and observe, this progress in one of mankind´s great adventures in science. And the search shall continue through new scientific landscapes for the benefit of the whole world.

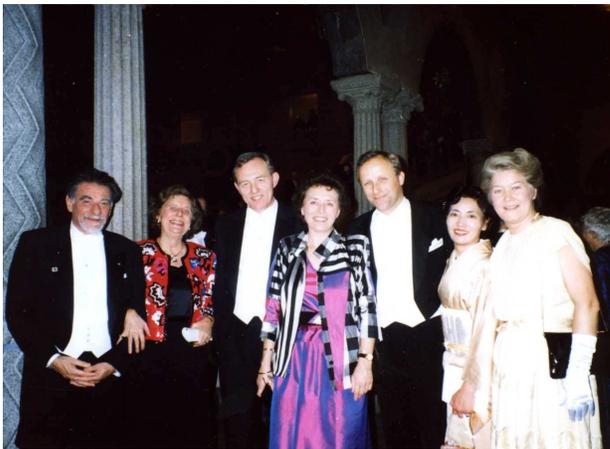

**Fig. 5.1** In Stockholm at the Nobel banquet in 1992: From the left, Alex, Inge, Kristian Fossheim, Elsa Fossheim, Oystein Fischer, Mrs Kitazawa, Hanne Fischer

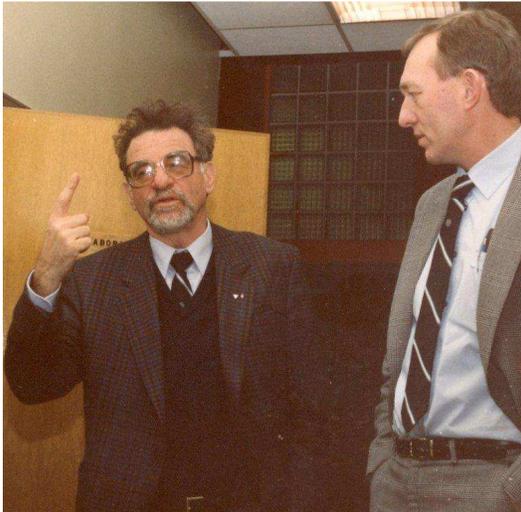

**Fig. 5.2** Alex and the author on Alex`s visit to Trondheim just after receiving the Nobel prize in 1987.

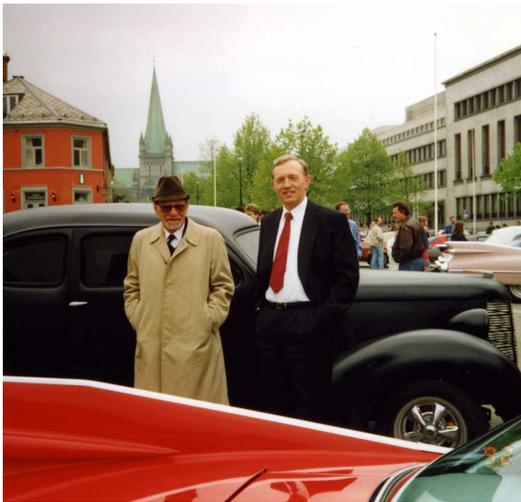

**Fig. 5.3** Alex and the author visiting an American car display in Trondheim, on the day of the honorary doctor degree ceremony in Trondheim in 1992

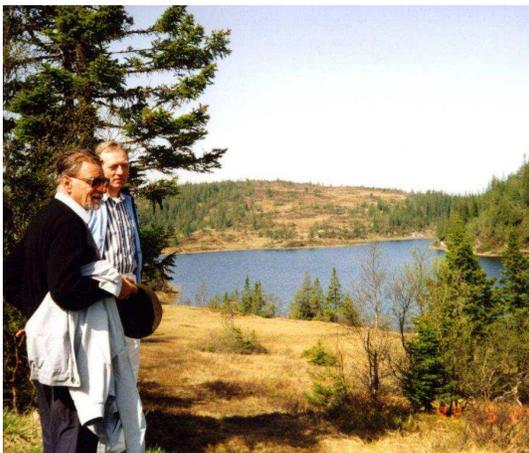

**Fig. 5.4** Alex and the author on an outing trip near Trondheim the day after he received his honorary doctor degree from University of Trondheim (later renamed the NTNU)

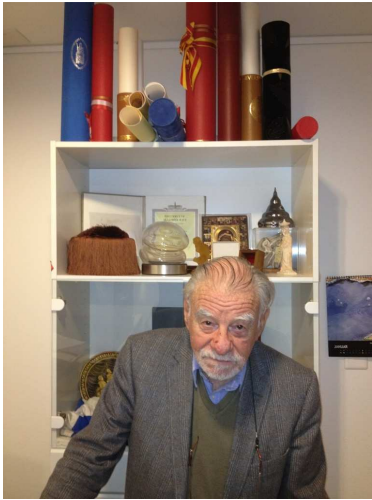
**Fig. 5.5** Alex in his Tertianum office in 2012